\documentclass[twocolumn,superscriptaddress,floatfix,preprintnumbers]{revtex4}
\usepackage{graphics,amssymb,amsmath,epsfig,color}
\usepackage{graphicx}

\begin{document}

\title{Universal Quantum Electron Microscopy:\\A Small-Scale Quantum Computing Application with Provable Advantage}
\author{Hiroshi Okamoto}
\email[]{okamoto@akita-pu.ac.jp}
\affiliation{Department of Intelligent Mechatronics, Akita Prefectural University, Yurihonjo 015-0055, Japan}
\date{\today}
\begin{abstract}
We propose a simple design of a quantum electron microscope that ``queries'' a beam-sensitive phase object, such as a biological specimen, as part of quantum computation. Lower quantum query complexity, not the time complexity, of a quantum algorithm means less specimen damage, which translates to more data extracted from the specimen. Hence small-scale quantum computing offers provable quantum advantage in this context. A possible application of the proposed microscope is the Grover search for a true structure, out of a set of candidate structures. 
\end{abstract}
\maketitle

Quantum query complexity is the number of calls a quantum computer (QC) needs to make to an ``oracle'' to solve a problem \cite{query_complexity}. The query model has been extensively studied in the field of quantum computing because it is relevant to many quantum algorithms and it also makes certain theoretical analyses tractable.

In a sense, the abstract concept of quantum query complexity becomes ``real'' in quantum measurement of fragile specimens. Specifically in the context of high-resolution biological electron microscopy (EM) \cite{cryoEM_textbook}, each query to the biological specimen, i.e., passing of a probe electron, damages the specimen with a certain probability. Hence, in principle, low query complexity of a quantum algorithm designed to obtain information about the ``oracle'', namely the specimen, translates to a measurement associated with a small amount of specimen damage. This in turn means that a large amount of information is obtainable from the specimen before we destroy it. Thus, query complexity is \emph{the} figure of merit of an algorithm designed for a given task in this context, rather than a proxy for more fundamental measures such as the time complexity. It is also worth noting the following: In the quantum computing community, certain algorithms such as Grover's algorithm have often been characterized as offering merely a ``modest'' polynomial speedup as opposed to an exponential speedup. In contrast, in the EM community people say ``every Angstrom counts'' with respect to resolution. Hence here is an opportunity for those modest algorithms to make a significant difference. A closely related point is that quantum advantage in our setting is free from subtleties that degrades practicality of those modest quantum algorithms in the purely computational setting  \cite{PRX_practicality_polynomial_Qadvantage}. 

An EM capable of querying the specimen in the above general sense has hardly been considered \cite{q_interface,resilient_QEM}. On the other hand, the use of quantum enhancement in EM, in order to image beam-sensitive specimens, has been discussed for more than a decade \cite{designs_QEM,Madan_review,TEM_q_limit} and also experimental results have begun to be reported \cite{Koppell_10kV,int_free_electrons}. Quantum-enhanced forms of EM are often referred to as quantum electron microscopy (QEM). Many, though not all, proposals of QEM aim at imaging weak phase objects beyond the shot noise limit to approach the Heisenberg limit \cite{eeem,multipass}. Note that biological specimens are usually regarded as weak phase objects in EM.

In this Letter, we present a \emph{universal} QEM design, which is able to perform anything programmable quantum mechanically. Universal QEM would make most sense when we perform tasks other than standard phase contrast imaging. We will find that those other tasks may include efficiently finding a known structure. Such measurements could indeed make sense in structural biology: Due to the recent developments in electron cryomicroscopy (cryoEM), now we are largely able to determine the atomic structure of a biological molecule by \emph{classical} averaging methods such as single particle analysis, provided that a large number of the molecule of interest are available \cite{single_particle_anal_review}. In contrast, quantum enhancement is required when only a single copy of the specimen is available. Tasks such as comprehensively identifying \emph{known} species of molecules in the crowded cellular environment, perhaps in the context of electron cryotomography \cite{template_matching}, may be a suitable arena for quantum technologies.

From the fundamental physics perspective, universal QEM definitely is possible. One could, in principle, connect an EM to a QC via suitable quantum interfaces \cite{optical_q_interface,q_interface} placed at both the illumination and detection sides of the EM. In this way, one could transfer, or teleport, a quantum state from the QC to the illuminating electron wave to the specimen and also transfer the state of the exit electron wave back to the QC. Let this be the \emph{definition} of what a universal QEM can do. The real question, on the other hand, is whether there exists a sufficiently simple and feasible scheme to do so. We answer this question in the affirmative, on the condition that the specimen is a pure phase object. In what follows, the symbol $e$ denotes the positron charge. Let the $z$-axis be the electron-optical axis. Let $\lambda$ be the wavelength of imaging electrons. A \emph{diffraction plane} is any plane conjugate to the back focal plane of the objective lens. An \emph{image plane} refers to any plane conjugate to the plane where the specimen is placed. We generally do not normalize a quantum state. 

\begin{figure}
\includegraphics[scale=0.3]{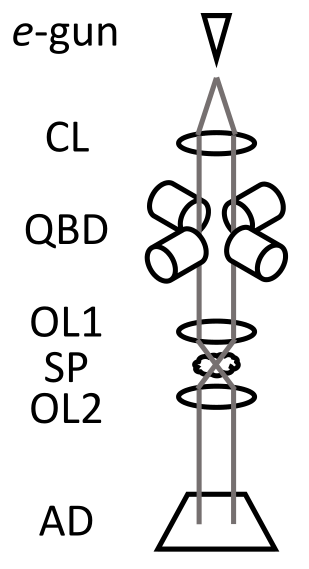}

\caption{Schematic drawing of a universal QEM at the conceptual level. It comprises a pulsed electron gun ($e$-gun), condenser lens (CL), quantum beam deflector (QBD), objective pre-field lens (OL1), specimen (SP), objective post-field lens (OL2), and a pixelated area detector (AD). Additional lenses, such as a projector lens, are not shown. To fully combat inelastic scattering events, energy of the scattered electron should be measured. \emph{In principle}, this can be done with time-of-flight.}
\end{figure}

Figure 1 shows our universal QEM scheme at the conceptual level. We defer discussions of physical realizations to later parts of this paper. At first glance, it is rather similar to the 4-dimensional (4D) scanning transmission EM (STEM) \cite{4D_STEM_review}. Following the electron gun and the condenser lens, there are two beam deflectors for bending the electron beam in the $x$ and $y$ directions at a diffraction plane. Below the pre- and post-field objective lenses and a specimen in between, a pixelated electron detector is placed at a diffraction plane. The crucial difference from the conventional STEM, however, is that the beam deflectors are quantum. Consider the beam deflector along the $x$-direction. This deflector \emph{is} a qudit, i.e., a $d$-level quantum system, with $d$ distinct quantum states $|0\rangle,|1\rangle,\cdots,|d-1\rangle$. These states are associated with, for example, a set of correspondingly equally-spaced amount of magnetic flux, which bends the electron beam. A superconducting flux qubit \cite{supercond_qubit_review}, for example, is a $d=2$ version of it. The same applies to the deflector in the $y$-direction. These two qudits, which we call qudit $x$ and qudit $y$, are part of a larger QC, equipped with as many additional qubits as necessary, that controls the QEM. Entanglement-assisted QEM \cite{eeem} may be regarded as the simplest version of the present scheme, with a single-axis deflector and a single-qubit QC. In EM, unlike standard quantum computation, a small-scale QC or even a single-qubit QC helps.

The effect of the quantum beam deflectors is the following. Reflecting the $d\times d$ distinct quantum states of the combined system of qudits $x$ and $y$, there correspond $d\times d$ points on the specimen, where the electron beam is focused. Hence one may raster-scan the electron beam by properly setting each qudit at proper times. What is newly enabled, however, is that one could also make a quantum superposition of various positions of the electron beam. Indeed, an arbitrary 2D structure of the electron beam may be generated, although it is not exactly an arbitrary structure of the electron wave front, because the electron state is heavily entangled with the two qudits. Nonetheless, it is rather remarkable that an arbitrary 2D structure can be generated by only two deflectors along the $x$ and $y$ directions. The ability of the two qudits to have entanglement between them enables this, and shows that a quantum instrument could, in a sense, occasionally be simpler than the classical counterpart.

Next, we show that our scheme is universal. We want to probe the phase shift of the phase object at the $d\times d$ locations quantum mechanically. Let these locations be indexed by two integers $p$ and $q$, where $0\le p<d$ and $0\le q<d$. Let the phase shift of the specimen at the location $\left(p,q\right)$ be $\theta_{p,q}$. For the universality defined earlier in this paper, the availability of the following operation, which we call an ``oracle call'', is sufficient:
\begin{equation}
|p,q\rangle\Rightarrow e^{i\theta_{p,q}}|p,q\rangle,\label{eq:oracle}
\end{equation}
where $|p,q\rangle$ is a quantum register in our QC, with $d^{2}$ states. (Here we somewhat enlarge the notion of the oracle from the one used in computer science, wherein $\theta_{p,q}$ is restricted to be either $0$ or $\pi$. Note that the latter can simulate the more widely used oracle that flips an ancilla qubit if and only if $\theta_{p,q}=\pi$.) The quantum register $|p,q\rangle$ turns out to be the combined qudits $x$ and $y$: We let $|p,q\rangle=|p\rangle\otimes|q\rangle$, where $|p\rangle$ and $|q\rangle$ are the states of qudits $x$ and $y$, respectively. To realize the transform of Eq. (\ref{eq:oracle}), we first produce an electron from the electron gun in the state $|0\rangle$, so that the initial state of the combined system of the electron and the beam deflector is $|0\rangle\otimes|p,q\rangle$. Let the electron state $|n,m\rangle$ be the one that is to be focused on the location $\left(n,m\right)$ of the specimen. The action of the beam deflector is, by definition, $|0\rangle\otimes|p,q\rangle\Rightarrow|p,q\rangle\otimes|p,q\rangle$. Next, we let the electron pass the specimen. By definition, we obtain $|p,q\rangle\otimes|p,q\rangle\Rightarrow e^{i\theta_{p,q}}|p,q\rangle\otimes|p,q\rangle$. Finally, we detect the electron in the far field. Since the electron wave from the point $\left(p,q\right)$ evolves into a plane wave in the far field, schematically we have 
\begin{equation}
|p,q\rangle=\int \frac{dk}{2\pi}\int \frac{dl}{2\pi}e^{i\left(kp+lq\right)}|k,l\rangle.\label{eq:e_diffraction}
\end{equation}
where $k$ and $l$ are real numbers that represents a point on the diffraction plane. Suppose that we detected the electron at the point $\left(k,l\right)$. This leaves the two qudits in the state
\begin{equation}
e^{i\theta_{p,q}}|p,q\rangle\otimes|p,q\rangle\Rightarrow e^{i\theta_{p,q}}\cdot e^{i\left(kp+lq\right)}|p,q\rangle.\label{eq:e_detection}
\end{equation}
Since we know $k$ and $l$ from our measurement, we can perform a phase shift operation to the qudits $x,y$ to obtain the state $e^{i\theta_{p,q}}|p,q\rangle$, which is the right-hand side of Eq. (\ref{eq:oracle}).

\begin{figure}
\includegraphics[scale=0.3]{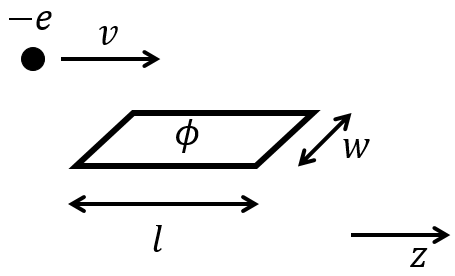}(a)
\includegraphics[scale=0.3]{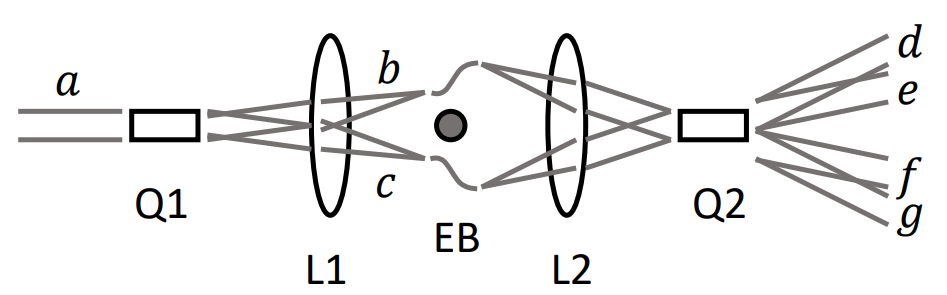}(b)

\caption{Designs of some parts of the universal QEM. (a) An electron with a velocity $v$ flies nearby a qubit with a trajectory parallel to the plane, on which the qubit is placed. The qubit, with the dimension $l\times w$, holds a magnetic flux $\phi$. (b) A qu\emph{d}it comprising multiple qubits. Two qubits (Q1, Q2) placed on diffraction planes deflect the electron beam minimally. The lens (L1) separates the deflected electron waves on the image plane, where a negatively biased electron biprism (EB) enlarges the separation between the two waves $b$ and $c$. Q1 deflects the incoming wave $a$ into waves $b$ and $c$ depending on its state, and Q2 deflects the waves $b$ and $c$ further into waves $f,g$ and $d,e$, respectively.}
\end{figure}

We proceed to discuss physical realizations of our scheme. The only nonstandard part in the scheme is the quantum beam deflector, and we focus on the part for the $x$-axis, i.e., the qudit $x$. While there may be other choices, below we consider superconducting quantum circuit as a physical platform because of its ability to produce a quantum-mechanically superposed electromagnetic field around it. In particular, the flux qubit can produce a superposition of two distinct amount of magnetic flux $\phi_{A}$ and $\phi_{B}$. As shown below, the magnetic flux difference $\Delta\phi=\phi_{A}-\phi_{B}$ has to be about the magnetic flux quantum $\phi_{0}=h/2e$ \cite{HO_Nagatani} or greater. Figure 2 (a) shows an electron with a velocity $v$ passing by a flux qubit, with the dimensions shown in the figure. The trajectory is bent depending on the qubit state. Since the electron wave has an angular spread $\approx\lambda/w$ after passing over the qubit due to diffraction, the angular deflection $\Delta p / p$ should satisfy
\begin{equation}
\frac{\Delta p}{p}=\frac{evB\Delta t}{p}=\frac{ev\phi}{plw}\cdot\frac{l}{v}=\frac{e\phi}{pw}>\frac{\lambda}{w}=\frac{h}{pw},
\end{equation}
where $p,\phi,B,\Delta t$ are, respectively, the momentum of the electron, the magnetic flux held by the qubit, the magnetic flux density and the time the electron takes to fly by the qubit. Hence we obtain the condition $\phi\gtrsim\phi_{0}$ to produce a quantum-mechanically distinct electron wave. Unfortunately, the magnetic flux held by a flux qubit is usually smaller than $\phi_{0}$ \cite{dwave_ferromag_qubits}. A conceptually simple way to work around this problem is to make a row of multiple flux qubits, each with states $|0\rangle$ and $|1\rangle$, and entangle all of them so that, as a whole, the entire set of qubits operates in the space spanned by the states $|00\cdots0\rangle$ and $|11\cdots1\rangle$. This entire set of qubits may then be regarded as a single qubit with $\Delta\phi\approx\phi_{0}$, which we call a full-vortex qubit (FVQ). They could be realized either by brute-force applications of quantum gates to entangle all the constituent qubits, or by designing certain interaction among the constituent qubits \cite{full_vortex_qubit}. One could then combine $d$ FVQs to realize the qudit, although we will describe another way with $\ln d$ scaling shortly.

Another possible way for realizing a FVQ is the use of the so-called bosonic qubit \cite{100_photon_qubit}. A bosonic qubit can store, in its microwave cavity, microwave photons in a quantum-mechanically controlled fashion. It was estimated \cite{okamoto_kaya_micron} that about $\alpha^{-1}$ photons, where $\alpha$ is the fine-structure constant, are needed to generate magnetic flux $\approx\phi_{0}$ in an instant when the photon energy is stored in the magnetic field. Since controlling $\approx100$ photons in a 3D cavity has been experimentally demonstrated \cite{100_photon_qubit}, we envision doing the same with a coplanar microwave cavity, above which an electron flies. On the other hand, back action to the qubit is estimated to be small \cite{suppl_mat_1}.

Figure 2 (b) shows a way to form a qudit using $\log_{2}d$ qubits. These qubits form a quantum register to represent an integer $0\le p<d$ in the binary form. Assuming that each qubit can only bend the electron trajectory by the angle $\delta\theta\approx\lambda/w$, we need to artificially enlarge the deflection angle except for the least significant qubit. To do so, we separate the waves $b,c$ from the qubit Q1 by letting them go through the lens L1 to arrive at an image plane. There we artificially enlarge the physical distance between the two waves by a classical means, for example by a negatively biased electron biprism. Other highly versatile methods for electron wave manipulation are also known \cite{electron_waveguide,Juffmann_smily}. After going through L2 to go to the next diffraction plane, the angle between the waves $b,c$ is enlarged to be $2\delta\theta$, and hence Q2 can split these waves into four waves $d,e,f$ and $g$. A similar argument applies to a system comprising more than two qubits. Finally, we note that the deflection angle is as small as $\delta\theta\approx 10^{-7}$ for $300\,\mathrm{keV}$ electrons and a typical dimension of microfabrication is $w \approx 10\,\mu\mathrm{m}$. Hence the enlargement of the deflection angle does not derail the electron wave off the qubit. 

The electron beam at each location $\left(p,q\right)$ on the specimen should be sufficiently focused. For our scheme to work, the diffracted beams from these locations should have similar intensity profiles, or otherwise the phase factor $e^{i\left(kp+lq\right)}$ in Eq. (\ref{eq:e_detection}) would be multiplied with an unwanted amplitude factor. However, high-angle elastic scattering up to $\approx10\,\mathrm{mrad}$ results from addition of scattered waves from relevant atoms with essentially random phase values. This entails rapidly fluctuating intensity in the diffraction plane \cite{meas_error}. To minimize detection of electrons in such a region, we enlarge the transmitted beam in the diffraction plane by sufficiently focusing the incident beam on the specimen. We estimate that this measure suffices to solve the problem \cite{suppl_mat_2}. 

Having discussed hardware designs, we proceed to consider software. In principle, any measurement method physically possible should be implementable, because of the universality of our scheme. In particular, multipass TEM \cite{multipass} is realized simply by repeated applications of Eq. (\ref{eq:oracle}), followed by a phase-contrast imaging steps involving a quantum Fourier transform, its inverse and a ``phase plate'' operation in between. In what follows, we discuss applications of Grover's algorithm to show more involved uses of universal QEM. We define two words to avoid confusion in the following discussions. We call the process described in Eq. (\ref{eq:oracle}) an \emph{oracle call}. We refer to a call from ``off-the-shelf'' Grover's algorithm, which expects a zero or $\pi$-phase shift, as a \emph{subroutine call}. 

The simplest, although artificial, application of Grover's algorithm is search for an object, or rather a single pixel A. Suppose that pixel A has phase shift $\pi$, while all other pixels have zero phase shift. In this case, Grover's algorithm enables us to find pixel A with $d$ queries, when there are $d^{2}$ pixels in total. If the phase shift is $\pi/k$ instead of $\pi$, then we can let $k$ electrons pass the specimen for each subroutine call from Grover's algorithm. This search makes sense if radiation damage to the specimen is delocalized and the entire specimen is destroyed after $n$ subroutine calls that satisfies $d\ll n\ll d^{2}$. On the other hand, this search does \emph{not} make sense if the specimen damage is localized to each pixel and we care about damage to the pixel A. Since the quantum amplitude at pixel A grows as $\approx\sin\left(\frac{\pi s}{2d}\right)$ at $s$-th iteration, the sum of the quantum probability at pixel A through $d$ iterations is approximately $\int_{0}^{d}\sin^{2}\left(\frac{\pi s}{2d}\right)ds=\frac{d}{2}.$ However, the sum of the probability at pixel A is $1$ if one ``classically'' measures the phase shift pixel by pixel against a reference. 

To go beyond the above contrived example, consider Grover search to find the right structure among $N$ candidate structures. Sequential testing would cost $\propto N$ queries, which we want to cut down to $\propto\sqrt{N}$. Let the $\alpha$-th candidate structure be associated with a phase map $\hat{\theta}^\alpha_{p,q}$. We let all these phase maps satisfy $\sum_{p,q}\hat{\theta}^\alpha_{p,q}=0$ without loss of generality. Let $\mathcal{P}_\alpha$  be a set consisting of pixels $\left(p,q\right)$ such that its size $|\mathcal{P}_\alpha|=d^2/2$ is half of all the pixels and it maximizes $\sum_{\left(p,q\right)\in \mathcal{P}_\alpha}\hat{\theta}^\alpha_{p,q}$. This tends to, but does not necessarily, make $\hat{\theta}^\alpha_{p,q}$ positive, where $\left(p,q\right) \in \mathcal{P}_\alpha$. Assume that the standard deviation of the phase map of the actual specimen $\theta_{p,q}$ is known to be of the order $\pi/k$. 

Our procedure to find the true structure is as follows. We begin with a superposition $\sum_{\alpha=1}^{N}|\alpha\rangle$ on a register of the QC and then produce $\sum_{\alpha=1}^{N}\bigl[|\alpha\rangle\otimes \sum_{\left(p,q\right)\in \mathcal{P}_{\alpha}}|p,q\rangle\bigr].$ After an oracle call, we obtain 
\begin{equation}
\sum_{\alpha=1}^{N}\Bigl[|\alpha\rangle\otimes \sum_{\left(p,q\right)\in \mathcal{P}_{\alpha}}e^{i\theta_{p,q}}|p,q\rangle\Bigr].
\end{equation}
Since we use a QC, we may freely rearrange the points $\left(p,q\right)\in \mathcal{P}_{\alpha}$ to obtain a linear configuration. Note that the gate count in the algorithm is not a primary concern to us. Specifically, we provide a bijection $f_\alpha$ from $\mathcal{P}_{\alpha}$ to $\mathcal{L}=\{1,2,\cdots,d^2/2\}$. Note that we have much freedom in choosing $f_\alpha$. Writing $\beta=f_\alpha(p,q)$ in each branch of the entire quantum state involving $|\alpha \rangle$, we obtain a state after the rearrangement 
\begin{equation}
\sum_{\alpha=1}^{N} \sum_{\beta=1}^{d^2/2} e^{i\Theta_{\alpha,\beta}}|\alpha\rangle\otimes|\beta\rangle ,\label{eq:before_beta_measurement}
\end{equation}
where $\Theta_{\alpha,\beta}$ represents permutated values of $\theta_{p,q}$ associated with $\mathcal{P}_{\alpha}$. To extract the mean value $\Theta_\alpha=\frac{2}{d^2}\sum_{\beta}\Theta_{\alpha,\beta}$, note that biological specimens are \emph{weak} phase object and we may write $e^{i\Theta_{\alpha,\beta}}\approx 1+i\Theta_{\alpha,\beta}$. Hence application of quantum Fourier transform (QFT) to Eq. (\ref{eq:before_beta_measurement}) with respect to $\beta$ yields an amplitude $\propto 1+i\Theta_\alpha$ at the zero-frequency state. Then we multiply $i$ to all the nonzero frequency states. This step, albeit for a different purpose, is reminiscent of the use of a $\pi/2$ phase plate in EM. We then apply inverse-QFT. These steps converts all the phase variation to amplitude variation. The resultant state is of the form
\begin{equation}
\sum_{\alpha=1}^{N} \sum_{\beta=1}^{d^2/2} e^{i\Theta_\alpha}(1+\eta_{\alpha,\beta})|\alpha\rangle\otimes|\beta\rangle ,\label{eq:before_beta_measurement2}
\end{equation}
Finally, assuming $\lvert\eta_{\alpha,\beta}\rvert \ll 1$, we measure $\beta$ to obtain $\approx\sum_{\alpha=1}^{N} e^{i\Theta_{\alpha}}|\alpha\rangle.\label{eq:after_beta_measurement}$ We expect $\Theta_\alpha \approx \pi/k$ for the correct hypothesis $\alpha$ because it should be about the standard deviation of the phase map. We expect much smaller $\Theta_\alpha$ for other $\alpha$. Hence we can compose a Grover subroutine call by repeating the process for $\approx k$ times. We have not analyzed the effect of various errors and our argument here should be regarded only as evidence that useful algorithms exist. 

Inelastic scattering ``mildly collapses'' the wavefunction in the real space to a finite area \cite{resilient_QEM, meas_error}. To be resistant to inelastic scattering, one may try to ensure the followings. Firstly, for any given point $(p,q)$ on the specimen, most $\mathcal{P}_\alpha$ should be designed to include a point that is close to $(p,q)$. Those $\mathcal{P}_\alpha$ that do not include such a point will lose the associated quantum amplitude upon inelastic scattering at $(p,q)$. Secondly, each $f_\alpha$ should be chosen, taking advantage of the large number of ways to do so, in such a way that measurement of $\beta$ after the inverse-QFT step does not eliminate surviving candidate structures. This roughly means that physically closer points $(p,q)$ should be converted to similar values of $\beta$ by the \emph{set} of $f_\alpha$. 

Final comments are in order. Firstly, there are quantum algorithms \cite{Kuperberg_algorithm, Montanaro_algorithm} that are able to efficiently find a user-specified structure and report its place. Although these algorithms work only for phase objects with phase shift values $0$ or $\pi$, they suggest the existence of useful quantum algorithms for QEM. Secondly, exponential quantum advantage of universal quantum measurement has been shown for some cases \cite{QLearning_experiments}, although its relevance to QEM is not yet clear at present. 

In summary, we have shown a simple universal QEM scheme. Also shown is evidence that useful quantum algorithms exist for QEM. Further study remains to be done. Useful algorithms should tolerate imprecise ``oracles'', inelastic scattering events, and preferably be executable on a noisy small-scale quantum computer. 

The author thanks Professor Robert M. Glaeser for discussions on the future of cryoEM. This research was supported in part by the JSPS ``Kakenhi'' Grant (Grant No. 19K05285).


%
%
\clearpage
\widetext 
\begin{center}
\textbf{\large Supplemental Material for ``Universal Quantum Electron Microscopy: \\
A Small-Scale Quantum Computing Application with Provable Advantage''}
\end{center}
\setcounter{equation}{0}
\setcounter{figure}{0}
\setcounter{table}{0}
\setcounter{page}{1}
\makeatletter
\renewcommand{\theequation}{S\arabic{equation}}
\renewcommand{\thefigure}{S\arabic{figure}}
\renewcommand{\bibnumfmt}[1]{[S#1]}
\renewcommand{\citenumfont}[1]{S#1}

In what follows, we use symbols defined in the main text unless noted otherwise. 

\section{Estimation of back action to superconducting qubits}

Here we roughly estimate the effect of flying electron on a superconducting qubit. We find that such ``back action" is small from the purely fundamental physics perspective: The qubit is ``protected'' by the factor of fine structure constant $\alpha \approx 1/137$ in a sense described below. However, proper engineering would be needed to ensure that the back action is indeed small. Henceforth we often ignore numerical factors of order $1$. We write Klitzing constant $R_K=h/e^2=25.8~\mathrm{k\Omega}$ and the impedance of the vacuum $Z_0 = \sqrt{\mu_0/\varepsilon_0}=377~\Omega$. Note the fact $Z_0/R_K=2\alpha$, where $\alpha$ is the fine structure constant $\alpha \approx 1/137$.

Before proceeding, few remarks are in order. First, interaction of an electron and a superconducting qubit is analogous to that of a controlled-not (CNOT) gate. Since CNOT gate is essentially symmetric, i.e. the control and target qubits swap their roles in the Hadamard-transformed basis, the ``back action" in this a sense is unavoidable. What we are attempting to show in this section is the ``control qubit", which is the superconducting qubit in the present case, in the original basis is unaffected by the interaction process to a good approximation. Second, the use of the Aharonov-Bohm (AB) effect would allow us to construct a superconducting qubit that is essentially free from back action in the above sense \cite{HO_Nagatani_suppl}. However, the engineering challenge associated with fabrication of such a qubit is rather significant. Third, some relevant calculation is presented in Ref. \cite{okamoto_kaya_micron_suppl}. 

We model the ``qubit'' as a lumped-circuit LC resonator, neglecting the effect of a Josephson junction. We have primarily the bosonic qubit in our mind, but the following argument is rather general and the essence of it should be valid more broadly. Our semiclassical analysis considers a classical electron that flies by the qubit. The Hamiltonian of the ``qubit'' is 
\begin{equation}
H = \frac{q^2}{2C} + \frac{\phi^2}{2L}, \label{eq:hamiltonian_suppl}
\end{equation}
where $C$, $L$, $q$, $\phi$ are the capacitance, inductance, the stored charge in the capacitor and the trapped magnetic flux in the inductor, respectively. Canonical quantization requires $[\phi,q]=i\hbar$. Following the standard procedure, we obtain a dimensionless Hamiltonian $h = H/\hbar\omega = Q^2/2 + \Phi^2/2$, where $\omega=1/\sqrt{LC}$, $Q=q/\sqrt{\hbar\omega C}$ and $\Phi=(\sqrt{\omega C/\hbar})\phi$. The dynamical variables $\Phi$ and $Q$ satisfy the commutation relation $[\Phi,Q]=i$, leading to the ladder operators $a=(\Phi + iQ)/\sqrt{2}$ and $a^\dag=(\Phi - iQ)/\sqrt{2}$.

Quantum charge fluctuation $\delta q$ in the circuit is as follows. The condition ${\delta q}^2/2C \approx \hbar \omega$ gives $\delta q \approx e \sqrt{R_K/Z_0} \approx e/\sqrt{\alpha}$ if $Z_0 \approx \sqrt{L/C}$. A formal calculation yields essentially the same, but weaker, result $\delta q = e\sqrt{R_K/4 \pi Z_0}$, again if we let $Z_0 = \sqrt{L/C}$. Nonetheless, this already gives us an indication that the effect of a flying electron to the circuit would be rather small, because the maximum charge an electron can induce on a capacitor plate is $e$, which is smaller than the quantum-mechanical charge fluctuation. Moreover, a symmetric capacitor design would reduce the induced charge significantly.

Next, consider quantum fluctuation $\delta \phi$ of the trapped magnetic flux. The condition ${\delta \phi}^2/2L \approx \hbar \omega$ gives $\delta \phi \approx \phi_0 \sqrt{Z_0/R_K} \approx \phi_0 \sqrt{\alpha}$, where $\phi_0 = h/e^2$ is the magnetic flux quantum. Hence in the magnetic case the quantum fluctuation is smaller than the natural unit of the quantity $\phi_0$ by the factor $\approx \sqrt{\alpha}$. On the other hand, the flying electron produces a magnetic flux density $B \approx \mu_0 I/l$, where $I$ is the current produced by the flying electron and $l$ is the distance between the electron and the LC resonator. For rough estimation purposes, we introduce precisely one characteristic length $l$ that describes the length scale of the circuit and electron-circuit interaction. Moreover, we simplify our analysis by regarding the velocity of the flying electron as $c = 1/\sqrt{\mu_0 \varepsilon_0}$, which indeed is not a bad approximation in TEM. Hence the current is $I \approx ec/l$, the associated magnetic flux density is $B \approx e Z_0/l^2$, and hence the magnetic flux produced by the electron, which is applied on the circuit, is $\phi \approx e Z_0 \approx \phi_0 \alpha$. Thus, despite the fact that quantum fluctuation is smaller than $\phi_0$ by the factor $\approx \sqrt{\alpha}$, the magnetic flux generated by the flying electron is smaller still by another factor $\approx \sqrt{\alpha}$. This suggests that the back action is insignificant in the magnetic case too. Moreover, again we should be able to further reduce the magnetic flux induced by the electron with a careful, perhaps symmetric, qubit design.

Next, we estimate the excitation probability of the LC resonator upon passing of the flying electron. Note that the bosonic qubit in particular uses basis states that do not have a definite number of photons and hence a single-photon excitation would not be fully destructive. We begin with the magnetic case. We regard passing of the flying electron as a small momentary external magnetic flux source and treat it as time-dependent perturbation. The perturbation potential to be added to the Hamiltonian of Eq. (\ref{eq:hamiltonian_suppl}) is 
\begin{equation}
V(t) = -\frac{\phi \phi_a}{L} = -\frac{\phi_a}{L}\sqrt{\frac{\hbar Z_0}{2}}(a+a^\dag),
\end{equation}
which lasts for a short time $\tau \approx l/c$ and makes the minimum point of the magnetic flux potential to be at the applied magnetic flux by the flying electron $\phi_a \approx \alpha \phi_0$. Time-dependent perturbation theory tells us that, at the first order, the excitation probability from the ground state $|0\rangle$ to the first excited state $|1\rangle$ is
\begin{equation}
p_{ex} = \frac{1}{\hbar^2}\left|\langle 1|V(t)|0\rangle\right|^2\left|\int_0^\tau e^{i\omega t}dt\right|^2. \label{eq:prob_excitation}
\end{equation}
Recalling $\phi_a \approx \alpha \phi_0$, we obtain 
\begin{equation}
\frac{1}{\hbar^2}\left|\langle 1|V(t)|0\rangle\right|^2 = \frac{\phi_a^2  Z_0}{2 \hbar L^2} \approx \frac{Z_0^3}{R_K L^2}
\end{equation}
The absolute square of the integration in Eq. (\ref{eq:prob_excitation}) is
\begin{equation}
\left|\int_0^\tau e^{i\omega t}dt\right|^2 = \frac{2}{\omega^2}(1-\cos{\omega t}) \approx t^2,
\end{equation}
where the last approximation is valid when $\omega t \ll 1$, which is not quite true in the present case, but the left hand side of the equation is smaller than $t^2$. Using the characteristic length $l$, we regard $t \approx l/c$ and $L \approx \mu_0 l$, which gives 
\begin{equation}
p_{ex} \approx \frac{Z_0}{R_K} \approx \alpha \approx 0.01.
\end{equation}
Our analysis is crude but rather robust, in the sense that Fermi's golden rule also gives basically the same result if we assume the density of state to be $\rho (E) \approx 1/(\hbar \omega)$ and regard 
\begin{equation}
\frac{1}{\omega} \approx \sqrt{LC} \approx \sqrt{\mu_0 l \varepsilon_0 l} \approx \frac{l}{c}.
\end{equation}

The analysis of the case of electrostatic excitation is similar to the magnetic case. The charge operator $q$ is expressed as 
\begin{equation}
q = \sqrt{\hbar \omega C}Q = \sqrt{\frac{\hbar}{Z_0}}Q = -i\sqrt{\frac{\hbar}{2Z_0}}(a-a^\dag).
\end{equation}
The perturbation potential is $V(q) = -q_a q/C$, where $q_a \approx e$ is the induced charge by the flying electron. Following a similar analysis, we again obtain the excitation probability of the order of $\alpha$.

\section{Uniformity of electron wave amplitude on the diffraction plane}

Diffracted beams from each point $(p,q)$ on the specimen should have an approximately identical intensity distribution on the diffraction plane for our scheme to work, as mentioned in the main text. What makes them vary is distinct atomic configurations associated with each ``point'' $(p,q)$, which result in various diffraction patterns. Our strategy is to use a sufficiently focused incident beam on the specimen, so that the smooth transmitted wave covers a large area of the diffraction plane to dominate the irregular diffraction pattern due to the scattered waves. However, the scattered waves would \emph{also} be enlarged, hence the necessity for a quantitative investigation. Below we exclusively consider $300~\mathrm{keV}$ probe electrons. We will often consider a coordinate system on the unit sphere, with the polar angle $\theta$ and the azimuthal angle $\varphi$ inherited from the spherical coordinate system.

First, we compute the elastic scattering probability for a typical biological specimen. We employ the expected number densities of biologically important elements (hydrogen, carbon, nitrogen, oxygen and sulfur) in a ``typical'' specimen, which have been computed in Appendix B of Ref. \cite{meas_error_suppl}. Let these number densities be $n_\mathrm{H}$, $n_\mathrm{C}$, $n_\mathrm{N}$, $n_\mathrm{O}$ and $n_\mathrm{S}$, respectively. These data are combined with the total scattering cross sections of the relevant atoms given in the NIST database \cite{NIST_database_suppl}. Hence we obtain the probability of elastic scattering for unit thickness of the specimen, which turns out to be $1.8\times10^{-3}/\mathrm{nm}$. For example, the probability of elastic scattering is $p_S=5.4~\%$ for a $30~\mathrm{nm}$-thick specimen, while it is $p_S=9.0~\%$ for a $50~\mathrm{nm}$-thick specimen. We will consider $p_S=5~\%$ and $10~\%$ probabilities below.

We assume that the angular intensity profile of the incident electron beam is gaussian. We write
\begin{equation}
I(\theta) = \frac{1}{2\pi\sigma^2} e^{-\frac{\theta^2}{2\sigma^2}},
\label{eq:incident_suppl}
\end{equation}
where $\theta$ is the convergence angle and $\sigma$ is a characteristic angle. We consider a typical value of $\sigma = 10~\mathrm{mrad}$ and a highly focused $50~\mathrm{mrad}$. The prefactor is set to satisfy
\begin{equation}
\int_0^{\infty}I(\theta) 2\pi\theta d\theta= 1,
\end{equation} 
where we approximated $\sin{\theta}$ as $\theta$, and the upper limit of integration $\pi$ is replaced with an infinity. Define intensity of the transmitted wave $T(\theta)=(1-p_S)I(\theta)$. 

Next, we consider the intensity $S(\theta)$ of scattered waves at a diffraction plane, generated by a \emph{plane} incident wave. The intensity $S(\theta)$ is directly related to differential scattering cross sections. We consider the function $S(\theta)$ that is averaged in the sense that interference fringes, caused by scattered waves from different atoms, are ignored. Put another way, we regard the phase values of the scattered waves to be random and take the average intensity. Differential scattering cross sections for the relevant elements are documented in the NIST database \cite{NIST_database_suppl}. The weighted average of these differential scattering cross sections, with $n_\mathrm{H}$, $n_\mathrm{C}$, $n_\mathrm{N}$, $n_\mathrm{O}$ and $n_\mathrm{S}$ as weights, directly gives $S(\theta)$ up to the overall normalization factor. We normalize $S(\theta)$ to satisfy
\begin{equation}
\int_0^{\pi}S(\theta) 2\pi\sin{\theta} d\theta= p_S.
\end{equation}

Before the main analysis, we note a property of the scattered wave originated from a plane incident wave. We consider weak phase objects, which is a valid assumption for real biological specimens. In this case the exit wave is described as $\psi (\boldsymbol{r}) = e^{i\xi (\boldsymbol{r})} \approx 1 + i\xi (\boldsymbol{r})$, where $\boldsymbol{r}$ is a position vector on the $xy$ plane in the real space and $\xi (\boldsymbol{r})$ is the phase map of the specimen. To get the wave function on the diffraction plane, we Fourier transform it to obtain 
\begin{equation}
\int \psi(\boldsymbol{r})e^{-i\boldsymbol{k}\cdot \boldsymbol{r}}d\boldsymbol{r}^2 = (2\pi)^2 \delta^2 (\boldsymbol{k}) + i\int \xi(\boldsymbol{r}) e^{-i\boldsymbol{k}\cdot \boldsymbol{r}}d\boldsymbol{r}^2 = (2\pi)^2 \delta^2 (\boldsymbol{k}) + i\Xi(\boldsymbol{k}).
\end{equation}
Notice $\Xi(\boldsymbol{-k}) = \Xi(\boldsymbol{k})^*$, since $\xi (\boldsymbol{r})$ is real. Equivalently, the probability amplitudes on the diffraction plane at $\boldsymbol{k}$ and $\boldsymbol{-k}$ are the negative of the complex conjugate of each other. For later purposes, henceforth we consider $\Xi$ as a function of the scattering angle $\theta$ and the azimuthal angle $\varphi$ rather than of the wave vector $\boldsymbol{k}$. The analogous relation $\Xi(\theta,\varphi+\pi) = \Xi(\theta,\varphi)^*$ should hold to a good approximation. Except for this rule, the scattered wave has a random phase value at each point. The absolute square $\left|\Xi(\theta,\varphi)\right|^2$, when adequately averaged to remove the ``interference fringes'', loses $\varphi$ dependence and equals $S(\theta)$ mentioned above.

Consider a focused incident beam with the intensity described in Eq.(\ref{eq:incident_suppl}) instead of a plane incident wave. Let the wave function of the incident wave simply be a real function $\psi_i(\theta) = \sqrt{I(\theta)}$ without any phase factor. The focused incident wave is nothing but a superposition of plane waves, with the ``weight'' $\psi_i(\theta)$. Thus, when an electron is detected at a point on the diffraction plane, the associated probability amplitude is a superposition of scattered waves, each originating from a plane wave, at various scattering angles with a gaussian weight $\psi_i(\theta)$. In other words, the probability amplitude of detecting a scattered electron is described by a convolution of a gaussian function $\psi_i$ and the scattered wave function $\Xi$ from a plane wave.

\begin{figure}
\includegraphics[scale=0.3]{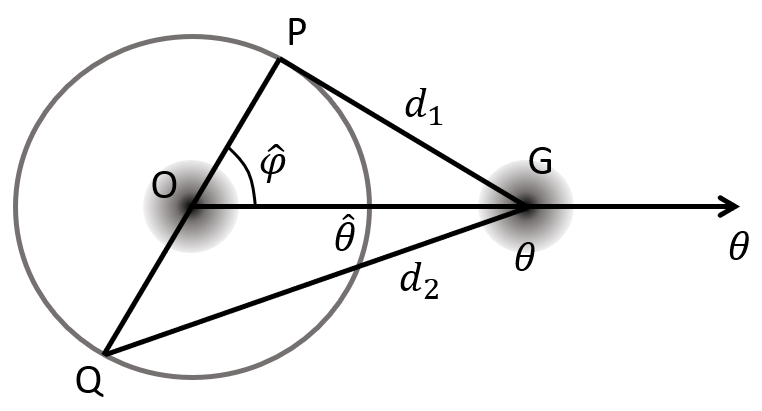}(a)\qquad
\includegraphics[scale=0.4]{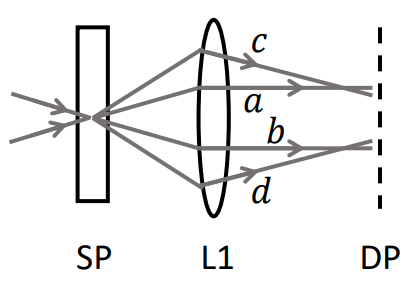}(b)
\caption{(a) Two-dimensional convolution on a unit sphere. The scattered wave distribution $\Xi$ is centered at the point O, while the gaussian incident wave $\psi_i$ is centered at the point G. The gray circle has the radius $\hat{\theta}$, on which the point P moves for integration. The variable $\hat{\theta}$ itself is also an integration variable. The distances $d_1$ and $d_2$, which are lengths PG and QG respectively, is computed using the spherical law of cosines. See the main text. (b) Specially designed lens for avoiding amplitude errors. Since the specimen (SP) is a phase object, the phase of the electron wave shifts upon transmission through the specimen (rays a and b). There may be scattered waves with large scattering angles (rays c and d) with non-smooth wave intensity on the diffraction plane (DP) after going through the lens (L1). This would adversely affect the measurement in our QEM scheme. We could employ a special lens L1 to merge rays c, d with rays a, b to avoid the adverse effect.}
\end{figure}

Figure S1 (a) shows exactly how the above-mentioned convolution is performed. Since our description is in terms of scattering angles, not the wave vector in the $xy$ plane, the diagram is on a unit sphere with the north pole O. The scattered wave amplitude distribution $\Xi(\theta,\varphi)$ is centered at O, while the gaussian function $\psi_i(\theta)$ is centered at the point G for the purpose of performing the convolution. Let the probability amplitude of detecting a scattered electron at the polar angle $\theta$ be $\psi_S(\theta)$. To perform convolution, we introduce a point P in Fig. S1 (a) that is specified by integration variables $\hat{\theta}$ and $\hat{\varphi}$. The convolution is given by 
\begin{equation}
\psi_S(\theta) = \frac{i}{\sqrt{\Delta\Omega}}\int_0^{\pi}\sin{\hat{\theta}}d\hat{\theta} \int_0^{2\pi}d\hat{\varphi} \,\Xi(\hat{\theta},\hat{\varphi})\psi_i(d_1),\label{eq:convolution_suppl}
\end{equation}
where $\Delta\Omega$ is a constant and $d_1$ is the distance between the points P and G on the unit sphere. The constant $\Delta\Omega$ is necessary for the following reason: Since $\left|\Xi(\theta,\varphi)\right|^2$ gives a probability when integrated over a solid angle, $\Xi(\theta,\varphi)$ has the ``dimension'' of the inverse of the square root of solid angle. Hence without the factor $(\Delta\Omega)^{-1/2}$, where $\Delta\Omega$ is a solid angle, the above integration would not be dimensionally consistent. More informally, while $\left|\Xi(\theta,\varphi)\right|^2$ is well-defined, $\Xi(\theta,\varphi)$ has random phase at \emph{each} point $(\theta,\varphi)$ and its magnitude depends on $\Delta \Omega$ as $\sqrt{\Delta \Omega}$, where $\Delta \Omega$ is the ``resolution'' of the solid angle that we are considering. This is because of a random-walk-like addition of probability amplitude within the small region $\Delta \Omega$. The necessity for this factor manifests also in later arguments. The distance $d_1$ is given by the spherical law of cosines as 
\begin{equation}
d_1=\cos^{-1}{\left(\cos{\theta}\cos{\hat{\theta}}+\sin{\theta}\sin{\hat{\theta}}\cos{\hat{\varphi}}\right)},
\end{equation}
where the range of the inverse cosine function is $[0,\pi]$. Since we know that $\Xi(\theta,\varphi+\pi) = \Xi(\theta,\varphi)^*$, we may rewrite the convolution as 
\begin{equation}
\psi_S(\theta) = \frac{i}{\sqrt{\Delta\Omega}}\int_0^{\pi}\sin{\hat{\theta}}d\hat{\theta} \int_0^{\pi}d\hat{\varphi} \,\left\{ \Xi(\hat{\theta},\hat{\varphi})\psi_i(d_1)  +\Xi(\hat{\theta},\hat{\varphi})^*\psi_i(d_2) \right\},\label{eq:convolution1_suppl}
\end{equation}
where $d_2$ is the distance between the points Q and G in Fig. S1 (a), which is given as 
\begin{equation}
d_2=\cos^{-1}{\left(\cos{\theta}\cos{\hat{\theta}}+\sin{\theta}\sin{\hat{\theta}}\cos\left({\hat{\varphi}+\pi}\right)\right)}.
\end{equation}
To further simplify Eq.(\ref{eq:convolution1_suppl}), we write $\Xi(\hat{\theta},\hat{\varphi}) = a(\hat{\theta},\hat{\varphi}) + ib(\hat{\theta},\hat{\varphi})$. This allows us to separate the real and imaginary part as $\psi_S(\theta) = i(A+iB)=-B+iA$, where 
\begin{equation}
B(\theta) = \frac{1}{\sqrt{\Delta\Omega}}\int_0^{\pi}\sin{\hat{\theta}}d\hat{\theta} \int_0^{\pi}d\hat{\varphi} \, b(\hat{\theta},\hat{\varphi})\left[\psi_i(d_1)  - \psi_i(d_2) \right]
\label{eq:convolution2_B_suppl}
\end{equation}
and 
\begin{equation}
A(\theta) = \frac{1}{\sqrt{\Delta\Omega}}\int_0^{\pi}\sin{\hat{\theta}}d\hat{\theta} \int_0^{\pi}d\hat{\varphi} \, a(\hat{\theta},\hat{\varphi})\left[\psi_i(d_1)  + \psi_i(d_2) \right].\label{eq:convolution2_A_suppl}
\end{equation}
The quantity $A(\theta)$ is the phase shift due to the specimen at around the focal point of the incident electron beam, and hence this represents what we want to measure. On the other hand, $B(\theta)$ is an undesirable amplitude to be added to the transmitted wave.

We are interested in the expected magnitude of the amplitude $B(\theta)$ when $b(\hat{\theta},\hat{\varphi})$ is regarded as random. Let $a(\hat{\theta})$, $b(\hat{\theta})$, $A(\theta)$ and $B(\theta)$ be random variables. It is natural to assume that $a(\hat{\theta})$ and $b(\hat{\theta})$ no longer have $\hat{\varphi}$ dependence. We assume that the collection of random variables $a(\hat{\theta})$ and $b(\hat{\theta})$ for each value of $\hat{\theta}$ are independent of each other. We put $a(\hat{\theta})$ and $b(\hat{\theta})$ on equal footing, and we also note that $\left|\Xi(\theta,\varphi)\right|^2 \approx S(\theta)$. Hence we may reasonably assume that $b(\hat{\theta})$ has the expected value $\overline{b(\hat{\theta})} = 0$ because of the random phase of $\Xi(\theta,\varphi)$, and its variance is $\mathrm{Var}(b(\hat{\theta})) = S(\hat{\theta})/2$. Since variance adds, $\mathrm{Var}\left(B(\theta)\right)$ can in principle be obtained simply as a linear combination of $\mathrm{Var}(b(\hat{\theta}))$. Then, the square root of $\mathrm{Var}\left(B(\theta)\right)$ should capture the notion of expected magnitude of amplitude $B(\theta)$. This undesirable amplitude is added to the transmitted wave, whose intensity is $T(\theta)$. Hence the relative amplitude error $E(\theta)$ when the electron is detected at the angle $\theta$, which is important in the QEM measurement scheme, may be expressed as  
\begin{equation}
E(\theta) = \sqrt{\frac{\mathrm{Var}\left(B(\theta)\right)}{\mathrm{Var}\left(B(\theta)\right) + T(\theta)}}.\label{eq:amplitude_error_suppl}
\end{equation}
The expected amplitude error $e_A$ averaged over all electron scattering angles is thus 
\begin{equation}
e_A = \int_0^{\pi} 2\pi \sin{\theta} d\theta E(\theta)\left\{ T(\theta) + S(\theta) \right\}.\label{eq:amplitude_error2_suppl}
\end{equation}

There is a subtlety associated with continuous integration when adding the variances. Recall that the variance of a random variable $Y = \sum_{k=1}^N c_k X_k$, which is a linear combination of a set of random variables $X_k$, is $\mathrm{Var}(Y) = \sum_{k=1}^N \sum_{l=1}^N c_k c_l \overline{X_k X_l}$. If all the random variables $X_k$ are independent to each other, then $\overline{X_k X_l} = \overline{X_k^2}\delta_{k,l}$. Hence we obtain $\mathrm{Var}(Y) = \sum_{k=1}^N c_k^2 \,\mathrm{Var}(X_k^2)$. We translate this discrete case to the case of continuous integration of random variables in Eq. (\ref{eq:convolution2_B_suppl}). For brevity, we rewrite Eq. (\ref{eq:convolution2_B_suppl}) as 
\begin{equation}
B(\theta) = \frac{1}{\sqrt{\Delta\Omega}}\int_{H}d\Omega \, b(\Omega)\left[\psi_i(d_1)  - \psi_i(d_2) \right],
\end{equation}
where $H$ is the unit hemisphere. The variance of $B(\theta)$ is 
\begin{equation}
\mathrm{Var}\left(B(\theta)\right) = \frac{1}{\Delta\Omega}\int_{H}d\Omega \int_{H}d\Omega' \, \overline{b(\Omega) b(\Omega')}\left[\psi_i(d_1)  - \psi_i(d_2) \right]\left[\psi_i(d_1')  - \psi_i(d_2') \right].\label{eq:variance_B}
\end{equation}
Let us discretize this integration to obtain a sum. As the discretized small solid angle, we take $\Delta \Omega$ introduced previously, a choice that will be justified shortly. Then we can replace the first integration by a sum as $(1/\Delta\Omega)\int_{H}d\Omega \Rightarrow \Sigma$, where the sum goes over all the small elements on the hemisphere. Next, we discretize the second integration as $\int_{H}d\Omega' \Rightarrow \Delta \Omega \Sigma$. Then the correlation function $\Delta\Omega' \, \overline{b(\Omega) b(\Omega')}$ has the dimension of the probability amplitude squared and we write 
\begin{equation}
\Delta\Omega' \, \overline{b(\Omega) b(\Omega')} = \overline{b(\Omega)^2}\,\Delta\Omega\,\delta_{\Omega,\Omega'},
\end{equation}
where $\delta_{\Omega,\Omega'}$ is the Kronecker delta that is $1$ when $\Omega$ and $\Omega'$ are the same small solid angle element, and $0$ otherwise. Hence Eq.(\ref{eq:variance_B}) is discretized as 
\begin{equation}
\mathrm{Var}\left(B(\theta)\right) = \sum\,\overline{b(\Omega)^2}\Delta\Omega\left[\psi_i(d_1)  - \psi_i(d_2) \right]^2 = \sum\,\frac{S(\Omega)}{2}\Delta\Omega\left[\psi_i(d_1)  - \psi_i(d_2) \right]^2,
\end{equation}
or back in the continuous and more explicit form 
\begin{equation}
\mathrm{Var}\left(B(\theta)\right) = \int_0^{\pi}\sin{\hat{\theta}}d\hat{\theta} \int_0^{\pi}d\hat{\varphi} \,\frac{S(\hat{\theta})}{2}\left[\psi_i(d_1)  - \psi_i(d_2) \right]^2.\label{eq:variance_B_suppl}
\end{equation}
The amplitude error $e_A$ is obtained from this result, Eq. (\ref{eq:amplitude_error_suppl}) and Eq. (\ref{eq:amplitude_error2_suppl}).

It remains to justify the identification of $\Delta \Omega$ in Eq. (\ref{eq:convolution_suppl}) and the solid angle resolution used in the above discretization. By definition, we have 
\begin{equation}
\left|\psi_S(\Omega)\right|^2 = \mathrm{Var}\left(A(\Omega)\right) + \mathrm{Var}\left(B(\Omega)\right) = \int_H d\hat{\Omega} \left\{ \mathrm{Var}(b(\hat{\Omega}))\left[\psi_i(d_1)  - \psi_i(d_2) \right]^2 + \mathrm{Var}(a(\hat{\Omega}))\left[\psi_i(d_1)  + \psi_i(d_2) \right]^2 \right\}.
\end{equation}
Noting the relation $\mathrm{Var}(b(\hat{\Omega})) = \mathrm{Var}(a(\hat{\Omega})) = S(\hat{\Omega})/2$, we obtain 
\begin{equation}
\left|\psi_S(\Omega)\right|^2 = \int_H S(\hat{\Omega})\left[\psi_i(d_1)^2  + \psi_i(d_2)^2 \right]d\hat{\Omega} = \int_S S(\hat{\Omega})\psi_i(d_1)^2 d\hat{\Omega} = \int_S S(\hat{\Omega})I(d(\Omega, \hat{\Omega}))d\hat{\Omega},\label{eq:intensity_convolution_suppl}
\end{equation}
where $d(\Omega, \hat{\Omega})$ is the distance between the two solid angle elements $\Omega$ and $\hat{\Omega}$ on the sphere. Note the change of the domain of integral from a hemisphere $H$ to the entire sphere $S$ in Eq. (\ref{eq:intensity_convolution_suppl}). Equation (\ref{eq:intensity_convolution_suppl}) says that convolution of the incident gaussian beam intensity, and the scattered wave intensity from a plane wave, equals the scattered wave intensity from the gaussian beam. Note that 
\begin{equation}
\int_S \left|\psi_S(\Omega)\right|^2 d\Omega = \int_S I(\Omega)d\Omega\int_S S(\Omega)d\Omega = p_S,
\end{equation}
as it should be. Thus, the disappearance of $\Delta \Omega$ in our final result is justified.

We numerically performed the above computation. About $\approx 600$ values of $\theta$ between $0$ and $\pi$ were used, which are identical to the $\theta$ values in the NIST database, wherein the density of data points is increasingly large near the origin $\theta = 0$. The angular resolution with respect to $\varphi$ was $\pi/100$ when the convolution was evaluated. 

Our results are as follows. Firstly, for $p_S=0.05$ representing a thin specimen, the expected amplitude error is $e_A = 3.7~\%$ for the highly focused incident beam with $\sigma = 50~\mathrm{mrad}$. The figure worsens to $e_A = 8.8~\%$ for a less focused beam with $\sigma = 10~\mathrm{mrad}$. Secondly, for a thick specimen with $p_S=0.1$, we obtained $e_A = 5.4~\%$ for $\sigma = 50~\mathrm{mrad}$ and $e_A = 13~\%$ for $\sigma = 10~\mathrm{mrad}$. For a QEM measurement scheme using $k$ electrons before measuring the qudits, these amplitude errors accumulate in the random-walk manner and are multiplied by a factor $\approx \sqrt{k}$. In the case of entanglement-enhanced electron microscopy scheme, which is the simplest QEM measurement procedure of the kind that we are discussing, a typical value of $k$ is about a few tens. Hence we may have to use a highly focused incident beam with $\sigma = 50~\mathrm{mrad}$, not to degrade the contrast by a factor $\approx \cos{\left(e_A\sqrt{8k}\right)}$ in the qubit measurement \cite{resilient_QEM_suppl}. 

A hardware solution may be needed in unusual experimental conditions. For example, we may have to use an electron beam with a small angular divergence because a highly focused electron beam would spread too much within a specimen if the thickness of the specimen is large. In this case, the above method may not work because the transmitted wave does not cover a large area in the diffraction plane. In such a case, dedicated electron optics could merge the nonuniform diffracted waves in the high scattering angle region with the relatively narrow transmitted wave, as shown in Fig S1 (b). Known versatile electron optical methods could be employed to realize it \cite{electron_waveguide_suppl, Juffmann_smily_suppl}. This solution fully obscures the fact that high-angle elastic scattering even happened. Note that, however, the merged waves coming from high-angle scattering still cause small amplitude and phase errors in the QEM measurement scheme.


\begin{thebibliography}{10}
\bibitem[1]{query_complexity} Peter Hoyer and Robert Spalek, Lower bounds on quantum query complexity, Bull. EATCS $\boldsymbol{87}$, 78-103 (2005). 

\bibitem[2]{cryoEM_textbook} R. M. Glaeser, K. Downing, D. DeRosier, W. Chiu, and J. Frank, \emph{Electron Crystallography of Biological Macromolecules} (Oxford University Press, New York, 2007).

\bibitem[3]{PRX_practicality_polynomial_Qadvantage} Ryan Babbush, Jarrod R. McClean, Michael Newman, Craig Gidney, Sergio Boixo, and Hartmut Neven, Focus beyond Quadratic Speedups for Error-Corrected Quantum Advantage, PRX Quantum $\boldsymbol{2}$, 010103 (2021).

\bibitem[4]{q_interface} H. Okamoto, Quantum interface to charged particles in a vacuum, Phys. Rev. A $\boldsymbol{92}$, 053805 (2015).

\bibitem[5]{resilient_QEM} H. Okamoto, Resilient quantum electron microscopy, Phys. Rev. A $\boldsymbol{106}$, 022605 (2022).

\bibitem[6]{designs_QEM} P. Kruit, R. G. Hobbs, C-S. Kim, Y. Yang, V. R. Manfrinato, J. Hammer, S. Thomas, P. Weber, B. Klopfer, C. Kohstall, T. Juffmann, M. A. Kasevich, P. Hommelhoff, and K. K. Berggren, Designs for a quantum electron microscope, Ultramicroscopy $\boldsymbol{164}$, 31-45 (2016).

\bibitem[7]{Madan_review} I. Madan, G. M. Vanacore, S. Gargiulo, T. LaGrange, and F. Carbone, The quantum future of microscopy: Wave function engineering of electrons, ions, and nuclei, Appl. Phys. Lett. $\boldsymbol{116}$, 230502 (2020).

\bibitem[8]{TEM_q_limit} Stewart A. Koppell, Yonatan Israel, Adam J. Bowman, Brannon B. Klopfer, and M. A. Kasevich, Transmission electron microscopy at the quantum limit, Appl. Phys. Lett. $\boldsymbol{120}$, 190502 (2022).

\bibitem[9]{Koppell_10kV} S. A. Koppell, M. Mankos, A. J. Bowman, Y. Israel, T. Juffmann, B. B. Klopfer, and M. A. Kasevich, Design for a 10 keV multi-pass transmission electron microscope, Ultramicroscopy $\boldsymbol{207}$, 112834 (2019).

\bibitem[10]{int_free_electrons} Amy E. Turner, Cameron W. Johnson, Pieter Kruit, and Benjamin J. McMorran, Interaction-Free Measurement with Electrons, Phys. Rev. Lett. $\boldsymbol{127}$, 110401 (2021).

\bibitem[11]{eeem} H. Okamoto, Possible use of a Cooper-pair box for low-dose electron microscopy, Phys. Rev. A $\boldsymbol{85}$, 043810 (2012).

\bibitem[12]{multipass} T. Juffmann, S. A. Koppell, B. B. Klopfer, C. Ophus, R. M. Glaeser, and M. A. Kasevich, Multi-pass transmission electron microscopy, Sci. Rep. $\boldsymbol{7}$, 1699 (2017).

\bibitem[13]{single_particle_anal_review} Dmitry Lyumkis, Challenges and opportunities in cryo-EM single-particle analysis, J. Biol. Chem. $\boldsymbol{294}$, 5181--5197 (2019).

\bibitem[14]{template_matching} Bronwyn A. Lucas, Benjamin A. Himes, Liang Xue, Timothy Grant, Julia Mahamid, and Nikolaus Grigorieff, Locating macromolecular assemblies in cells by 2D template matching with cisTEM, eLife 10:e68946 (2021).

\bibitem[15]{optical_q_interface} O. Kfir, Entanglements of Electrons and Cavity Photons in the Strong-Coupling Regime, Phys. Rev. Lett. $\boldsymbol{123}$, 103602 (2019).

\bibitem[16]{4D_STEM_review} Colin Ophus, Four-Dimensional Scanning Transmission Electron Microscopy (4D-STEM): From Scanning Nanodiffraction to Ptychography and Beyond, Microsc. Microanal. $\boldsymbol{25}$, 563--582 (2019).

\bibitem[17]{supercond_qubit_review} Morten Kjaergaard, Mollie E. Schwartz, Jochen Braumueller, Philip Krantz, Joel I.J. Wang, Simon Gustavsson, and William D. Oliver, Superconducting Qubits: Current State of Play, Annu. Rev. Condens. Matter Phys. $\boldsymbol{11}$, 369-395 (2020).

\bibitem[18]{HO_Nagatani} Hiroshi Okamoto and Yukinori Nagatani, Entanglement-assisted electron microscopy based on a flux qubit, Appl. Phys. Lett. $\boldsymbol{104}$, 062604 (2014).

\bibitem[19]{dwave_ferromag_qubits} T. Lanting, A. J. Przybysz, A. Yu. Smirnov, F. M. Spedalieri, M. H. Amin, A. J. Berkley, R. Harris. F. Altomare, S. Boixo, P. Bunyk, N. Dickson, C. Enderud, J. P. Hilton, E. Hoskinson, M. W. Johnson, E. Ladizinsky, N. Ladizinsky, R. Neufeld, T. Oh, I. Perminov, C. Rich, M. C. Thom, E. Tolkacheva, S. Uchaikin, A. B. Wilson, and G. Rose, Entanglement in a Quantum Annealing Processor, Phys. Rev. X $\boldsymbol{4}$, 021041 (2014).

\bibitem[20]{full_vortex_qubit} Hiroshi Okamoto, Full-vortex flux qubit for charged-particle optics, Phys. Rev. A $\boldsymbol{97}$, 042342 (2018).

\bibitem[21]{100_photon_qubit} Brian Vlastakis, Gerhard Kirchmair, Zaki Leghtas, Simon E. Nigg, Luigi Frunzio, S. M. Girvin, Mazyar Mirrahimi, M. H. Devoret , R. J. Schoelkopf, Deterministically Encoding Quantum Information Using 100-Photon Schroedinger Cat States, Science $\boldsymbol{342}$, 607-610 (2013).

\bibitem[22]{okamoto_kaya_micron} Hiroshi Okamoto, Reza Firouzmand, Ryosuke Miyamura, Vahid Sazgari, Shun Okumura, Shota Uchita, Ismet I. Kaya, TEM at millikelvin temperatures: Observing and utilizing superconducting qubits, Micron $\boldsymbol{161}$, 103330 (2022).

\bibitem[23]{suppl_mat_1} See Supplemental Material I.

\bibitem[24]{electron_waveguide} Robert Zimmermann, Michael Seidling, and Peter Hommelhoff, Charged particle guiding and beam splitting with auto-ponderomotive potentials on a chip, Nat. Commun. $\boldsymbol{12}$, 390 (2021).

\bibitem[25]{Juffmann_smily} Marius Constantin Chirita Mihaila, Philipp Weber, Matthias Schneller, Lucas Grandits, Stefan Nimmrichter, and Thomas Juffmann, Transverse electron-beam shaping with light, Phys. Rev. X, $\boldsymbol{12}$, 031043 (2022).

\bibitem[26]{meas_error} Hiroshi Okamoto, Measurement errors in entanglement-assisted electron microscopy, Phys. Rev. A $\boldsymbol{89}$, 063828 (2014).

\bibitem[27]{suppl_mat_2} See Supplemental Material II.

\bibitem[28]{Kuperberg_algorithm} G. Kuperberg, A subexponential-time quantum algorithm for the dihedral hidden subgroup problem, SIAM J. Comput. $\boldsymbol{35}$, 170--188 (2005).

\bibitem[29]{Montanaro_algorithm} Ashley Montanaro, Quantum Pattern Matching Fast on Average, Algorithmica $\boldsymbol{77}$, 16--39 (2017).

\bibitem[30]{QLearning_experiments} Hsin-Yuan Huang, Michael Broughton, Jordan Cotler, Sitan Chen, Jerry Li, Masoud Mohseni, Hartmut Neven, Ryan Babbush, Richard Kueng, John Preskill, and Jarrod R. McClean, Quantum advantage in learning from experiments, Science $\boldsymbol{376}$, 1182--1186 (2022).
\end{thebibliography}

\begin{thebibliography}{1}
\bibitem[1]{HO_Nagatani_suppl} Hiroshi Okamoto and Yukinori Nagatani, Entanglement-assisted electron microscopy based on a flux qubit, Appl. Phys. Lett. $\boldsymbol{104}$, 062604 (2014).

\bibitem[2]{okamoto_kaya_micron_suppl} Hiroshi Okamoto, Reza Firouzmand, Ryosuke Miyamura, Vahid Sazgari, Shun Okumura, Shota Uchita, Ismet I. Kaya, TEM at millikelvin temperatures: Observing and utilizing superconducting qubits, Micron $\boldsymbol{161}$, 103330 (2022).

\bibitem[3]{meas_error_suppl} Hiroshi Okamoto, Measurement errors in entanglement-assisted electron microscopy, Phys. Rev. A $\boldsymbol{89}$, 063828 (2014).

\bibitem[4]{NIST_database_suppl} A. Jablonski, F. Salvat, C. J. Powell, and A. Y. Lee , NIST Electron Elastic-Scattering Cross-Section Database Version 4.0, NIST Standard Reference Database Number 64, National Institute of Standards and Technology, Gaithersburg MD, 20899 (2016); https://srdata.nist.gov/srd64/, (retrieved Oct. 22, 2023).

\bibitem[5]{resilient_QEM_suppl} Hiroshi Okamoto, Resilient quantum electron microscopy, Phys. Rev. A $\boldsymbol{106}$, 022605 (2022). See Sec. IV-H.

\bibitem[6]{electron_waveguide_suppl} Robert Zimmermann, Michael Seidling, and Peter Hommelhoff, Charged particle guiding and beam splitting with auto-ponderomotive potentials on a chip, Nat. Commun. $\boldsymbol{12}$, 390 (2021).

\bibitem[7]{Juffmann_smily_suppl} Marius Constantin Chirita Mihaila, Philipp Weber, Matthias Schneller, Lucas Grandits, Stefan Nimmrichter, and Thomas Juffmann, Transverse electron-beam shaping with light, Phys. Rev. X, $\boldsymbol{12}$, 031043 (2022).
\end{thebibliography}
\end{document}